# Mathematical modelling and study of the encoding readout scheme for position sensitive detectors

Xiaoguang Yue [a,b], Ming Zeng [a,b,*], Zhi Zeng [a,b], Yi Wang [a,b], Xuewu Wang [a,b], Ziran Zhao [a,b], Jianping Cheng [a,b] and Kejun Kang [a,b]

[a] *Key Laboratory of Particle & Radiation Imaging (Tsinghua University), Ministry of Education, China*

[b] *Department of Engineering Physics, Tsinghua University, Beijing 100084, China*

**Abstract**

Encoding readout methods based on different schemes have been successfully developed and tested with different types of position-sensitive detectors with strip-readout structures. However, how to construct an encoding scheme in a more general and systematic way is still under study. In this paper, we present a graph model for the encoding scheme. With this model, encoding schemes can be studied in a more systematic way. It is shown that by using an encoding readout method, a maximum of $n(n-1)/2+1$ strips can be processed with n channels if n is odd, while a maximum of $n(n-2)/2+2$ strips can be processed with n channels if n is even. Furthermore, based on the model, the encoding scheme construction problem can be translated into a problem in graph theory, the aim of which is to construct an Eulerian trail such that the length of the shortest subcycle is as long as possible. A more general approach to constructing the encoding scheme is found by solving the associated mathematical problem. In addition, an encoding scheme prototype has been constructed, and verified with MRPC detectors.

*Keywords*: encoding readout method, mathematical modelling, graph theory, Eulerian trail, ERLP

## 1. Introduction

In many applications, the precise position of a charged particle in a relatively large detection area is desired. An example is the cosmic muon tracker detectors in a muon tomography facility. Because of the low flux of cosmic muons at sea level, an encoded readout method, rather than the conventional individual channel readout, can be used to reduce the size of the readout electronics.

In [1], we described an encoding readout method based on the fine-fine anode-array configuration, which has been used in multi-anode microchannel array (MAMA) detectors on board the Hubble Space Telescope (HST)[2]. In this method, $n(n+2)$ strips are divided into $2n+2$ groups, where $n$ is even, and strips in the same group are connected to the same electronics channel. An example scheme based on this configuration is shown in Figure 1. The encoding scheme uniquely encodes every two adjacent strips $\{i, i+1\}_{strip}$ to an unordered pair of electronics channels $\{a, b\}_{channel}$, making it possible to decode the particle position without ambiguity. The method performs well with multi-gap resistive plate chamber (MRPC) detectors at our muon tomography facility. Moreover, we proposed an approach to recovering the original charge distribution (see Section 2), which makes the readout method, for some extent, lossless.

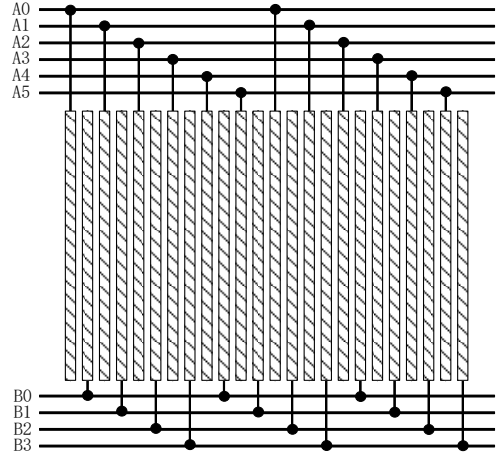

Figure 1. An encoding scheme based on the fine-fine configuration

Recently, Procureur et al. presented a similar method — "genetic multiplexing" [3], which was designed and tested with a Micromegas prototype. Moreover, a practical approach to constructing an encoding scheme with a prime number of channels was also proposed. The encoding scheme can be described as a channel number list showing the connections between strips and channels in strip order. Furthermore, this construction can be extended with minor adaptations to the cases where p is not prime by taking the first available channel number that forms a new doublet with the previous number.

Table 1. The number list of 7 channels constructed with the approach proposed in [3]. The connections between strips and channels are described in the form *Channel$_{strip}$*.

| | | | | | | |
|---|---|---|---|---|---|---|
| $1_1$ | $2_2$ | $3_3$ | $4_4$ | $5_5$ | $6_6$ | $7_7$ |
| $1_8$ | $3_9$ | $5_{10}$ | $7_{11}$ | $2_{12}$ | $4_{13}$ | $6_{14}$ |
| $1_{15}$ | $4_{16}$ | $7_{17}$ | $3_{18}$ | $6_{19}$ | $2_{20}$ | $5_{21}$ |
| $1_{22}$ | | | | | | |

The encoding readout method appears to be a general channel reduction technique for position sensitive detectors with strip-readout structure. The encoding scheme is the most important part of the method. However, although two encoding scheme construction approaches are available, how to construct an encoding scheme in a more general and systematic way remains unknown.

In this paper, we present a graph model for the encoding scheme. With the model, encoding schemes can be studied in a systematic way. It is shown that when using an encoding readout method, a maximum of $n(n-1)/2 + 1$ strips can be processed with n channels if n is odd, while a maximum of $n(n-2)/2 + 2$ strips can be processed with n channels if n is even. Furthermore, based on the model, the encoding scheme construction problem can be translated into a problem in graph theory, in which the aim is to construct an Eulerian trail such that the length of the shortest subcycle is as long as possible. A general approach to constructing the encoding scheme is found by solving the associated mathematical problem.

## 2. Statement of the problem

The encoding scheme is capable of recovering the charge distribution, which can be achieved in two steps (an example can be found in [1] and Section 6):

1. Decode the position of the two strips around the avalanche centre, which always carry more charge than other strips.
2. Reconstruct the charge distribution of the event on the basis of the predefined connections between the strips and the electronics channels.

Accordingly, two constraints should be considered during the construction of the encoding scheme to guarantee that the charge distribution can be reconstructed:

A. Any unordered pair of channels should be used at most once in the scheme.
   Otherwise, position ambiguity may occur in decoding step 1;

B. The minimum distance between two strips connected to the same read-out channel should be as large as possible.
   Because the charge distribution of an event can be reconstructed only if the fired strips are connected to different channels, the charge distributions of the events with a cluster size smaller than the minimum distance are guaranteed to be constructed. In other words, the minimum distance between two strips connected to the same electronics channel determines the upper bound of the guaranteed reconstructable charge distribution range (GRCDR) of the scheme. As shown in Figure 2, if the charge distribution of an event exceeds the GRCDR, the charge distribution of the event may not be reconstructed.

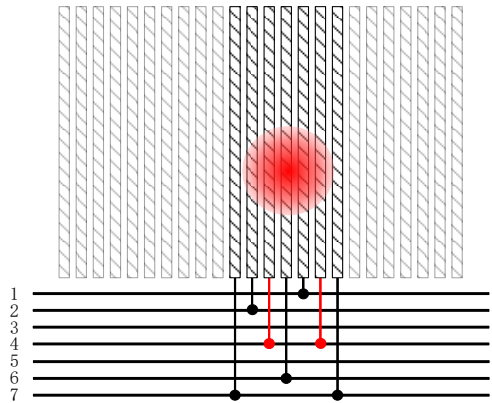

Figure 2.   An event with a charge distribution exceeding the GRCDR

## 3. Mathematical model

Firstly, an encoding scheme can be represented by a channel number list. The $i$th element in the list is the channel number to which the $i$th strip is connected. Assume a graph in which the vertices denote the electronics channels and the edges joining two vertices denote the unordered pairs of the two associated channels. Then, the channel list that describes the connections of an encoding scheme can be translated into a walk in the graph. Furthermore, because the unordered pairs of channels are not repeatable, the walk does not visit the same edge twice, which is called a trail in graph theory. For example, for the above-mentioned encoding schemes, the fine-fine configuration and genetic multiplexing can be represented by graph models, as illustrated in Figure 3 and Figure 4.

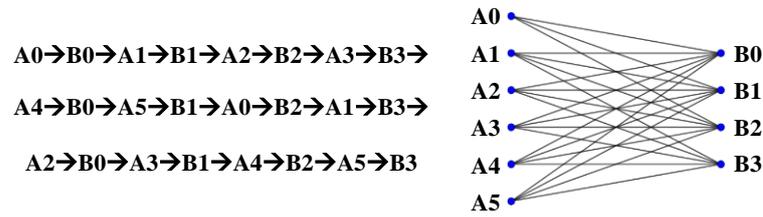

Figure 3.  The graph model of the encoding scheme shown in Figure 1 (the fine-fine configuration)

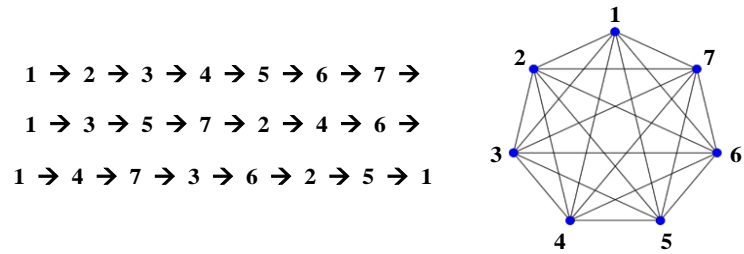

Figure 4.  The graph model of the encoding scheme described in Table 1 (the genetic multiplexing)

Conversely, an encoding scheme satisfying constraint A can be constructed by constructing a trail (a walk in which all the edges are distinct), which can be achieved by constructing an Eulerian trail. An Eulerian trail is a trail that traverses every edge in a graph exactly once. If the trail starts and ends at the same vertex (in other words, it is a closed Eulerian trail), then it is called an Eulerian circuit (or Eulerian tour). And a graph is Eulerian if it contains an Eulerian circuit.

Moreover, in the model, the cases where two strips are connected to the same channels are translated into the subcycles in the trail. Thus, constraint B implies that the length of the shortest subcycle in the trail should be as long as possible.

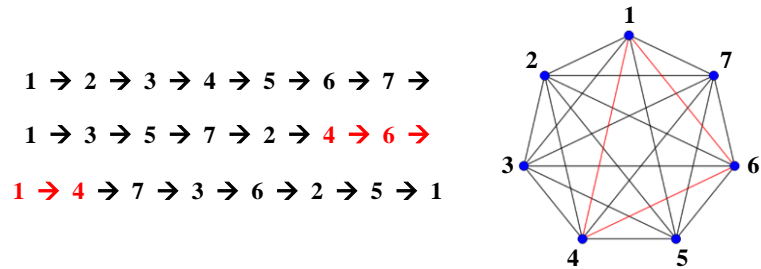

Figure 5.  The case where two strips are connected to the same channels can be translated into a subcycle in the trail

After mathematical modelling, the construction of an encoding scheme is converted into a problem in graph theory, in which the aim is to construct an Eulerian trail such that the length of the shortest subcycle is as long as possible.

## 4.  Maximum number of strips which can be processed with *n* channels

It is obvious that the number of strips encoded in a scheme is related to the number of unordered pairs used in the channel number list. As there are $n(n-1)/2$ unique unordered pairs with $n$ electronics channels, $n(n-1)/2 + 1$ strips is the upper bound of the number of strips which can be processed with n electronics channels by using the encoding readout method.

Using the mathematical approach, a more precise result can be given. Because the number of unordered pairs presented in the number list is equal to the length of the equivalent trail, which is defined as the number of edges in the trail, the problem can be translated to the calculation of the maximum length of an Eulerian trail with n vertices. This question is highly related to the existence of Eulerian trails in a graph, which can be checked with the Theorem and the Corollary [4] shown below:

**Theorem**

A nonempty connected graph[1] is Eulerian if and only if it has no vertices of odd degree[2].

**Corollary**

A connected graph has an Euler trail if and only if it has at most two vertices of odd degree.

Based on the theorem and the corollary, the maximum length of an Eulerian trail can be calculated:

i) The maximum length of an Eulerian trail with an odd number of vertices

A graph with an odd number of vertices and all unique edges connecting two vertices is an odd order complete graph. Because the vertices are all of even degree, these graphs are Eulerian and contain an Eulerian circuit. Therefore, the maximum length of the Eulerian trail with an odd number of vertices is $n(n-1)/2$, which reaches the upper bound.

ii) The maximum length of an Eulerian trail with an even number of vertices

A graph with an even number of vertices and all unique edges connecting two vertices is an even order complete graph. Because the degree of all vertices in these graphs is even, these graphs contain no Eulerian trail. According to the Corollary, at least *n*/2-1 edges must be deleted to make the graphs have exactly two odd vertices. Therefore, the maximum length of the Eulerian trail with an even number of vertices is $n(n-2)/2 + 1$, which is $n/2 - 1$ smaller than the upper bound.

As a result, the maximum number of strips that can be processed with n channels is

$$\text{maximum number of strips} = \begin{cases} \dfrac{n(n-1)}{2} + 1 & \text{if } n \text{ is odd} \\ \dfrac{n(n-2)}{2} + 2 & \text{if } n \text{ is even} \end{cases}$$

## 5. A novel approach to constructing an encoding scheme

As stated above, the encoding problem can be converted to the construction of an Eulerian trail in which the length of the shortest subcycle is as long as possible. A similar problem in the graph field is the Eulerian Recurrent Length Problem (ERLP)[5], which is intended to find an Eulerian circuit such that the length of the shortest subcycle is as long as possible; the Eulerian Recurrent Length (ERL) is defined as "the maximum of the shortest cycle length of Eulerian circuits". In [5],

---

[1] A graph which is connected in the sense of a topological space, i.e., there is a path from any point to any other point in the graph. A graph that is not connected is said to be disconnected.
[2] The degree of a vertex is defined as the number of edges connecting this vertex.

Jimbo discussed the Eulerian recurrent lengths of complete graphs and provided an approach to constructing an Eulerian circuit with an odd number of vertices, in which the shortest length of subcycles is n-4. Based on this result, a new approach to constructing an encoding scheme, which is named "Eulerian Trail Approach", can be proposed.

i) Method for an odd number of electronics channels.

Assume that the vertex set denoting the electronics channels is $\{0,1,2,\cdots,n-1\}$.

1. Construct the Hamiltonian path[3]
$$H_k = n - 1 \to v_0(k) \to v_1(k) \to \cdots \to v_{n-2}(k)$$
where $v_i(k)$ is defined as follows:

$$v_i(k) = \begin{cases} k & \text{if } i = 0 \\ \left(k + \dfrac{i+1}{2}\right) \bmod (n-1) & \text{if } i > 0 \text{ and } i \text{ is odd} \\ \left(k - \dfrac{i}{2}\right) \bmod (n-1) & \text{otherwise} \end{cases}$$

2. Let the Eulerian trail $P_n$ be $P_n = H_0 \to H_1 \to \cdots \to H_{(n-3)/2} \to n - 1$

The length of the Eulerian trail constructed is $n(n-1)/2$, which reaches the upper bound; and the distance between the occurrences of the same vertex in the trail is at least *n*-2 (see Appendix B for proof).

For instance, the Eulerian trail of 9 channels can be constructed as follows:

$$8 \to 0 \to 1 \to 7 \to 2 \to 6 \to 3 \to 5 \to 4 \to$$
$$8 \to 1 \to 2 \to 0 \to 3 \to 7 \to 4 \to 6 \to 5 \to$$
$$8 \to 2 \to 3 \to 1 \to 4 \to 0 \to 5 \to 7 \to 6 \to$$
$$8 \to 3 \to 4 \to 2 \to 5 \to 1 \to 6 \to 0 \to 7 \to 8$$

In the corresponding encoding scheme, the number of strips processed is 37; and the minimum distance between two strips connected to the same electronics channel is 7.

ii) Method for an even number of electronics channels.

Assume that the vertex set denoting the electronics channels is $\{0,1,2,\cdots,n-1\}$.
1. Construct the Eulerian trail of vertex set $\{0,1,2,\cdots,n-2\}$ with the method for an odd number of vertices.
2. Insert *n* at the position between the $n/2$th and the $(n/2+1)$th element in each Hamilton path and add an *n* at the end of the Eulerian trail.

The length of the Eulerian trail constructed is $n(n-2)/2 + 1$, which reaches the upper bound as well; and the distance between the occurrences of the same vertex in the trail is at least *n*-4.

For instance, an Eulerian trail of 10 channels can be constructed as follows:

$$8 \to 0 \to 1 \to 7 \to 2 \to 9 \to 6 \to 3 \to 5 \to 4 \to$$
$$8 \to 1 \to 2 \to 0 \to 3 \to 9 \to 7 \to 4 \to 6 \to 5 \to$$
$$8 \to 2 \to 3 \to 1 \to 4 \to 9 \to 0 \to 5 \to 7 \to 6 \to$$
$$8 \to 3 \to 4 \to 2 \to 5 \to 9 \to 1 \to 6 \to 0 \to 7 \to 8 \to 9$$

---
[3] Hamiltonian path is a path that visits each vertex exactly once.

In the corresponding encoding scheme, the number of strips processed is 42; and the minimum distance between two strips connected to the same electronics channel is 6.

## 6. Verification and testing

6.1. Experimental setup

An encoding scheme, which encodes 105 strips to 15 channels as shown in Figure 7, was constructed with the novel "Eulerian Trail Approach", and verified with MRPC detectors. The experimental setup of verification is shown in Figure 6, the output signal from the strips of the detectors were connected to the encoding board, and the encoded signals are connected to the readout electronics and data acquisition (DAQ) system. The cables used for connection in this experiment were HHSC series Ribbon Coax Cables from Samtec. A muon telescope system consisting of two plastic scintillators located above and below the MRPC detectors was used to generate trigger signals for the DAQ system.

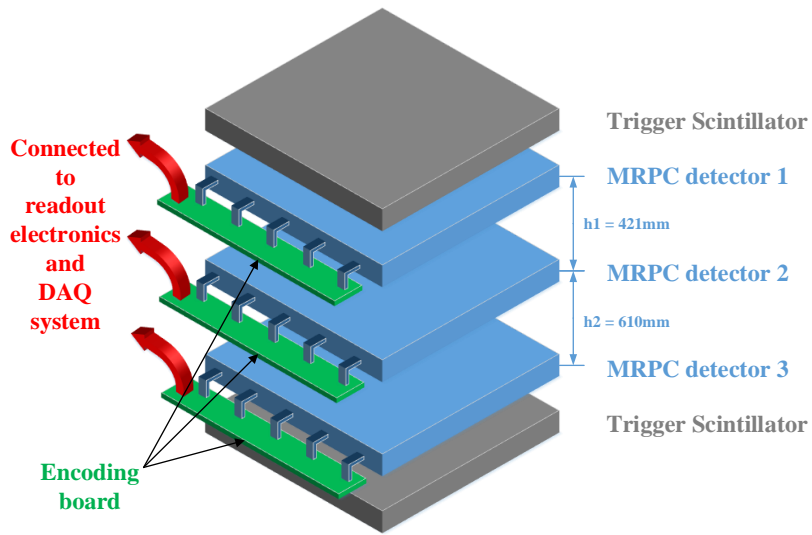

Figure 6. The experimental setup.

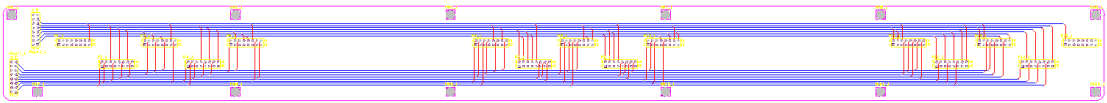

Figure 7. The scheme of the encoding board used in the experiment

The structure of the MRPC detector used is the same as described in [1]. Inside the detector, the readout strips are connected alternatively to the two ends as shown in Figure 8. In this experiment, to avoid the usage of long cables, only the strips at one end were connected to the encoding board and the strips at the other end were left floating and not read out. Therefore, the equivalent strip pitch in this experiment was 6mm, which was twice of the actual stripe pitch 3mm. The working gas inside the MRPC was a mixture of R134A/Isobutane/SF6 (90/5/5), and the working high voltage applied was +/- 7500V.

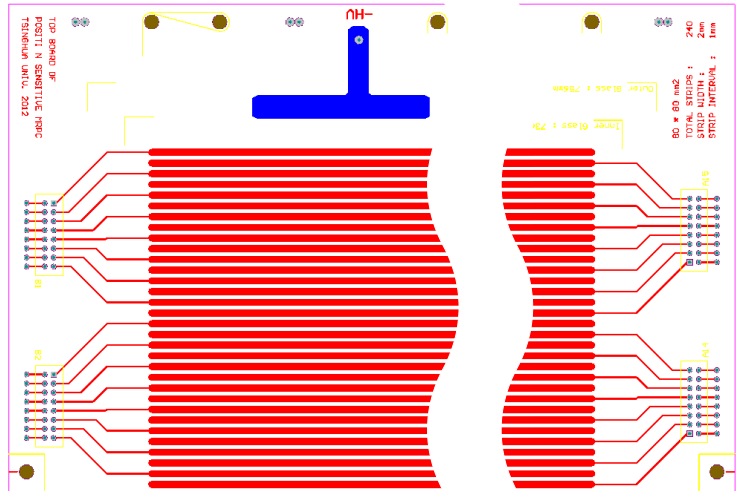

Figure 8.  A part of the readout strip scheme in the MRPC detector

The readout electronics and data acquisition (DAQ) system used in this experiment was the charge measurement system designed for BESIII drift chamber [6], which was based on flash ADCs with digital pipelines. Upon receiving a trigger, the digitized pulse shape data was delivered to an on-line PC via Ethernet.

6.2. Result

The data was analysed offline based on the ROOT data analysis frame [7]. The threshold discriminating signal from noise was set at $3 \times$ RMS above the signal baseline, and the charge value carried in a signal was extracted through numerical integration.

The 15 channels of encoding signals of a typical event are shown in Figure 9. In this particular event, there are four channels above the signal threshold, namely 2,8,9 and 14.

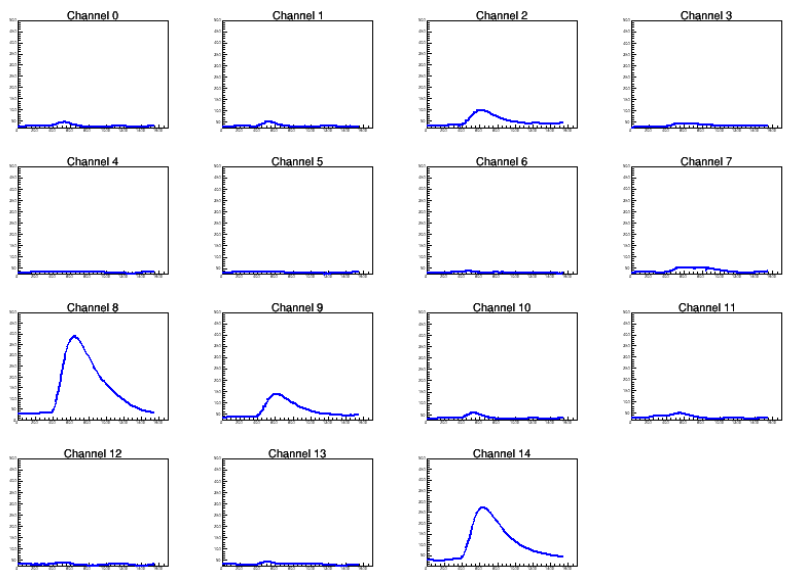

Figure 9.  Encoded signals from the encoding channels of a typical event

The pulses of these four channels are picked out and shown in Figure 10(a). In this event, the two channels which carry more charge are channel 8 and channel 14. Because there is only one adjacent strip pair $\{28,29\}_{strip}$ (the strip index starts from 0) connected to the two channels $\{8,14\}_{channel}$ in the encoding scheme, the cluster center can be located between strip 29 and strip 30 at first. After that, based on the connections in the encoding scheme, $28_{strip} \leftrightarrow 9_{channel}$, $29_{strip} \leftrightarrow 8_{channel}$, $30_{strip} \leftrightarrow 14_{channel}$ and $31_{strip} \leftrightarrow 2_{channel}$, the charge distribution of the event can be reconstructed as shown in Figure 10(b).

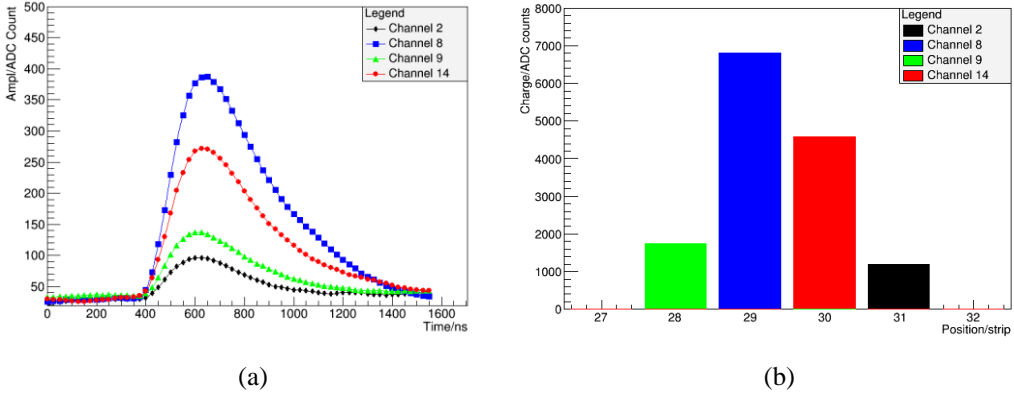

(a)                  (b)

Figure 10. (a)    The output signals above the threshold

(b)    The reconstructed charge distribution after decoding

With the reconstructed charge distribution, the event position can be reconstructed by calculating the charge centroid. The reconstructed position distribution along a MRPC detector is shown in Figure 11. The distribution is uniform as expected, except the deficit on the right due to the misalignment between the detector and the muon telescope.

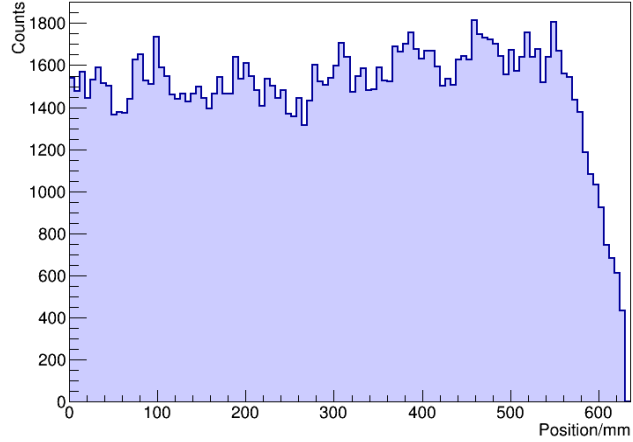

Figure 11.    The reconstructed position distribution of a detector

To further verify the readout scheme, the position resolution of the detectors is also measured with a similar method used in [8]. The residual $x_{res}$ is defined as the difference between the measured position $x_2$ and the linear interpolation position from $x_1$ and $x_3$.

$$x_{res} = x_2 - \left(\frac{h_2}{h_1 + h_2}x_1 + \frac{h_1}{h_1 + h_2}x_3\right)$$

Assume that the position resolutions of the three detectors are equal and denoted by $\sigma(x)$, the

standard deviation of the residual distribution

$$\sigma(x_{res}) = \sqrt{1 + \frac{h_1^2 + h_2^2}{(h_1 + h_2)^2}} \sigma(x)$$

And thus, the position resolution of the detectors

$$\sigma(x) = \sigma(x_{res}) \bigg/ \sqrt{1 + \frac{h_1^2 + h_2^2}{(h_1 + h_2)^2}}$$

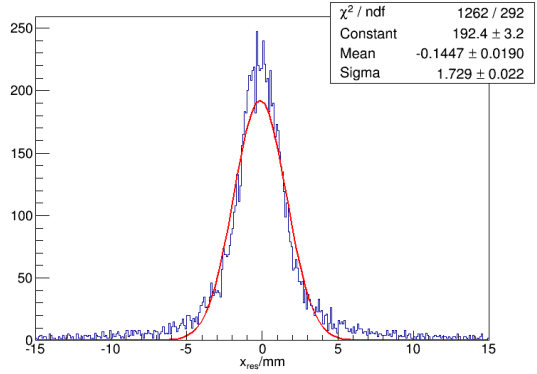

Figure 12.　The distribution of the position residual with Gaussian fit result

In this experiment, the standard deviation of the residual distribution is $\sigma(x_{res}) = 1.729$ mm as shown in Figure 12. Therefore, the position resolution of the detectors $\sigma(x) = 1.404$ mm after correction. This result is reasonable considering the 6 mm equivalent strip pitch in the experiment.

## 7. Conclusion

In this paper the problem of the systematic definition of an optimal encoding scheme for the read-out of *m* strips with *n* electronics channel (*m* > *n*) was investigated and validated. We found the maximum number of strips that can be processed with *n* electronics channels. Moreover, we translated the problem of the definition to a problem in graph theory, i.e. the construction of an Eulerian trail having the maximum possible length of its shortest subcycle. Using this novel approach, we presented a general method to construct the optimal encoding scheme. In addition, an encoding scheme prototype has been constructed and verified with MRPC detectors.

However, position ambiguities may arise in some special cases. For instance, the decoding method presented in this paper is based on the assumption that the two adjacent strips around the avalanche center carry more charge than the other strips. But this assumption may fail in some special cases, such as when a particle crosses the detector with a relatively large angle. Another case is when two or more particles penetrate the sensitive area within the acquisition window. For muon tomography the event rate is relatively low, but when the detector becomes larger for large object tomography, the possibility of multi-particle events will be significant and should be taken into account. In the near future, we will investigate and try to propose a practical encoding readout method for muon tomography facility. At the same time, the degradation of position resolution due to the encoding readout method will also be evaluated.

# Acknowledgements

This work is supported by the National Natural Science Foundation of China (No.11322548 and No.11305093) and Tsinghua University Initiative Scientific Research Program (No.2014Z21016).

**Appendix A.    A glossary of mathematical terms appeared in this paper [4]**

**Graph, vertices and edges**

A *graph* $G$ is an ordered triple $(V(G), E(G), \psi_G)$ consisting of a nonempty set $V(G)$ of *vertices*, a set $E(G)$, disjoint from $V(G)$, of *edges*, and an *incidence function* $\psi_G$ that associates with each edge of $G$ an unordered pair of (not necessarily distinct) *vertices* of $G$. If $e$ is an edge and $u$ and $v$ are vertices such that $\psi_G(e) = uv$, then $e$ is said to *join* $u$ and $v$; the vertices $u$ and $v$ are called the *ends* of $v$. (p1)

**Simple graph**

An edge with identical ends is called a *loop*. A graph is *simple* if it has no loops and no two of its links join the same pair of vertices. (p3)

**Complete graph**

A simple graph in which each pair of distinct vertices is joined by an edge is called a complete graph. Up to isomorphism, there is just one complete graph on *n* vertices; it is denoted by $K_n$. (p4)

**Connected graph**

A graph $H$ is a *subgraph* of $G$ (written $H \subseteq G$) if $V(H) \subseteq V(G)$, $E(H) \subseteq E(G)$, and $\psi_H$ is the restriction of $\psi_G$ to $E(H)$. (p8)

Two vertices $u$ and $v$ are said to be *connected* if there is a $(u,v)$-path in $G$. Connection is an equivalence relation on the vertex set $V$. Thus there is a partition of $V$ into nonempty subsets $V_1, V_2, \ldots, V_\omega$ such that two vertices $u$ and $v$ are connected if and only if both $u$ and $v$ belong to the same set $V_i$. The subgraphs $G[V_1], G[V_2], \ldots, G[V_\omega]$ are called the *components* of $G$. If $G$ has exactly one component, $G$ is *connected*; otherwise $G$ is *disconnected*. (p13)

**Degree**

The ends of an edge are said to be *incident* with the edge, and vice versa. (p3)

The *degree* $d_G(v)$ of a vertex $v$ in $G$ is the number of edges of $G$ incident with $v$, each loop counting as two edges. (p10)

**Walk, trail, path and the lengths of them**

We use the symbols $v(G)$ and $\varepsilon(G)$ to denote the numbers of vertices and edges in graph $G$. (p3)

A *walk* in $G$ is a finite non-null sequence $W = v_0 e_1 v_1 e_2 v_2 \ldots e_k v_k$, whose terms are alternately vertices and edges, such that, for $1 \leq i \leq k$, the ends of $e_i$ are $v_{i-1}$ and $v_i$. (p12)

The integer $k$ is the *length* of $W$. (p12)

If the edges $e_1, e_2, \ldots, e_k$ of a walk $W$ are distinct, $W$ is called a *trail*; in this case the length of $W$ is just $\varepsilon(W)$. If, in addition, the vertices $v_0, v_1, \ldots, v_k$ are distinct, $W$ is called a *path*. (p12)

**Cycle**

A walk is *closed* if it has positive length and its origin and terminus are the same. A closed trail whose origin and internal vertices are distinct is a *cycle*. (p14)

**Euler trail and Euler tour**

A trail that traverses every edge of $G$ is called an *Euler trail* of $G$ because Euler was the first to investigate the existence of such trails in graphs. (p51)

A *tour* of $G$ is a closed walk that traverses each edges of $G$ at least once. An *Euler tour* is a tour which traverses each edge exactly once (in other words, a closed Euler trail). A graph is *eulerian* if it contains an Euler tour. (p51)

**Hamilton path and Hamilton cycle**

A path that contains every vertex of $G$ is called a *Hamilton path* of $G$; similarly, a *Hamilton cycle* of $G$ is a cycle that contains every vertex of $G$.(p53)

**Appendix B.   The distance between the occurrences of the same vertex in the trail constructed with the method for an odd number of vertices is at least *n*-2**

**Proof.**

In the trail, the following hold.

$v_i(k) = v_i(k+1)$ implies $n + j - i \geq n - 2$ for each $k \in \{0, 1, \ldots, (n-5/2)\}$.

Because

Case 1: $i = 0, \; j = 0$.

Then, $k = k + 1$.

This case is impossible.

Case 2: $i = 0, \; j > 0$ and $j$ is odd.

Then, $k = \left(k + 1 + \frac{j+1}{2}\right) \mod (n-1)$, which implies $1 + \frac{j+1}{2} \equiv 0 \mod (n-1)$.

This case is impossible.

Case 3: $i = 0, \; j > 0$ and $j$ is even.

Then, $k = \left(k + 1 - \frac{j}{2}\right) \mod (n-1)$, which implies $j = 2$.

In this case, $+j - i = n + 2$.

Case 4: $i > 0$ and $j$ is odd, $j = 0$.

Then, $\left(k + \frac{i+1}{2}\right) \mod (n-1) = k + 1$, which implies $i = 1$.

In this case, $+j - i = n - 1$.

Case 5: $i > 0$ and $j$ is odd, $j > 0$ and $j$ is odd.

Then, $\left(k + \frac{i+1}{2}\right) \mod (n-1) = \left(k + 1 + \frac{j+1}{2}\right) \mod (n-1)$, which implies

$j - i = -2$.

In this case, $+j - i = n - 2$.

Case 6: $i > 0$ and $j$ is odd, $j > 0$ and $j$ is even.

Then, $\left(k + \frac{i+1}{2}\right) \mod (n-1) = \left(k + 1 - \frac{j}{2}\right) \mod (n-1)$, which implies

$\frac{i+j-1}{2} \equiv 0 \mod (n-1)$.

This case is impossible.

Case 7: $i > 0$ and $j$ is even, $j = 0$.

Then, $\left(k - \frac{i}{2}\right) \mod (n-1) = k + 1$, which implies $\frac{i}{2} + 1 \equiv 0 \mod (n-1)$.

This case is impossible.

Case 8: $i > 0$ and $j$ is even, $j > 0$ and $j$ is odd.

Then, $\left(k - \frac{i}{2}\right) \mod (n-1) = \left(k + 1 + \frac{j+1}{2}\right) \mod (n-1)$, which implies

$i = n - 2$ and $j = n - 3$.

In this case, $+j - i = n - 1$.

Case 9: $i > 0$ and $j$ is even, $j > 0$ and $j$ is even.

Then, $\left(k - \frac{i}{2}\right) \mod (n-1) = \left(k + 1 - \frac{j}{2}\right) \mod (n-1)$, which implies

$j - i = -2$.

In this case, $+j - i = n - 2$.